\documentclass[12pt]{article}
\usepackage{amsmath,amssymb}
\newtheorem{theorem}{Theorem}

\newtheorem{exe}[theorem]{Exercise}
\newtheorem{exa}[theorem]{Example}

\newtheorem{remark}[theorem]{Remark}

\newcommand{\CM}{{\cal M}}
\newcommand{\CQ}{{\cal Q}}

\def\la{\lambda}        
         
\def\om{\omega}

\newcommand{\rref}[1]{(\ref{#1})} 

\newcommand{\del}{{\partial}}
\def\parpo#1#2{\{#1,#2\}}

\def\HJ{Hamilton--Jacobi}

\def\var{manifold}
\def\bih{bi-Ham\-il\-tonian}
\def\bil{bi-Lagrangian}

\def\varb{\bih\ \var}

\def\ham{Hamiltonian}
\def\vefi{vector field}
\def\parp{Poisson bracket}

\def\parpu{{\parpo{\cdot}{\cdot}}}

\def\secoo{separation coordinates}
\def\coos{coordinates}
\def\neusys{Neumann system}
\def\LC{Levi-Civita}
\begin{document}
\begin{center}
{\Large \bf Bi-Hamiltonian aspects of the separability\\ of the
Neumann system}
\end{center}
\begin{center}
{Marco Pedroni}\\ \bigskip
Dipartimento di Matematica, Universit\`a di Genova\\
Via Dodecaneso 35, I-16146 Genova, Italy\\
E--mail: pedroni@dima.unige.it
\end{center}
\begin{abstract}
The Neumann system on the 2-dimensional sphere is used as a tool to
convey some ideas on the bi-Hamiltonian point of view on separation of
variables. It is shown that, from this standpoint, its separation
coordinates and its integrals of motion can be found in a systematic
way.
\end{abstract}

\section{Introduction}

Separation of variables for the \HJ\ equation is a classical topic
that is still very much investigated and has connections with several
important research fields, such as stationary reductions of soliton
equations (see \cite{DKN} and references therein), algebraic
completely integrable systems \cite{AHH93,Hur},
Riemannian geometry \cite{Ben,Kalninsbook,Kiyohara,Woodhouse},
B\"acklund transformations and Baxter's $Q$-operator
\cite{Sklyaninlectures,KuVa}, and string 
theory \cite{GoNeRu}.

The aim of this paper is to present, in the simple example of the
Neumann system, the main ideas of a new approach 
\cite{FMP2,cetraro,FP}
to separation of variables. This point of view is based on the
geometry of \varb s, and has been successfully applied to the
stationary reductions of the KdV hierarchy \cite{FMPZ2} and to 
Toda lattices \cite{creta}. 

The Neumann system is a well-known and very much studied mechanical
system (see, e.g., \cite{Moser,Audin,Harnadlectures,ratiu}),
given by a (mass 1) particle moving on the (unit) sphere under
the influence of a quadratic potential $V(x,y,z)=\frac12(a_1
x^2+a_2 y^2+a_3 z^3)$, where $a_1<a_2<a_3$ and $(x,y,z)$ are
Cartesian coordinates whose origin coincides with the center of
the sphere. In 1859, Carl Neumann showed that the \HJ\ equation
of this system is additively separable in the so-called
spheroconical (or elliptical spherical) coordinates
$(\la_1,\la_2)$, given by
\[
\frac{x^2}{\la-a_1}+\frac{y^2}{\la-a_2}+\frac{z^2}{\la-a_3}=
\frac{(\la-\la_1)(\la-\la_2)}{(\la-a_1)(\la-a_2)(\la-a_3)}\ ,
\]
with $a_1<\la_1<a_2<\la_2<a_3$. Once they have been introduced, one
can check that they are separation coordinates, but the
problem we want to address is to find out these coordinates in a
systematic way. It is known that the separation variables of the
Neumann system can be supplied by a Lax representation, but then the
problem is to find such a representation for a given system, which is
in general a quite difficult task. We will show that a careful study
of the Neumann system will allow us to deduce in a natural way the
spheroconical coordinates as 
separation variables. 

The plan of the paper is the following. In Section 2 we will introduce
the idea (due to A. Nijenhuis) of the geometrization of a coordinate
system by means of a suitable tensor field $L$. This allows one to
find a kind of compatibility condition between $L$ and a given
Hamiltonian
$H$, ensuring that the coordinates induced by $L$ separate the
\HJ\ equation associated with $H$. This condition is an intrinsic
(i.e., coordinate-free) form of the classical \LC\ separability
conditions. Before applying this results
to the
\neusys\ in Section 4, we will comment on their meaning in the theory
of \varb s in Section 3. Section 5 is devoted to two
additional results on the \neusys, that can be easily obtained from our
standpoint (using some facts highlighted in \cite{IMM}): A
simple rule for the construction of the integrals of motion and
the extension of the \bih\ structure from the cotangent bundle
$T^*S^2$ of the sphere to a 5-dimensional manifold, in order to
give the \neusys\ a \bih\ formulation, which is missing in
$T^*S^2$. The final section contains some concluding remarks.
\par\smallskip\noindent
{\bf Acknowledgments.}
The results presented in this paper have been obtained in
collaboration with Gregorio Falqui and Franco Magri, to whom I am very 
grateful. I wish to thank 
also Claudio Bartocci, Sergio Benenti, and Giovanni Rastelli for
useful discussions, and the organizers of NEEDS 2001 for the
opportunity to present these results there and for the nice atmosphere
at the conference. This work has been partially supported by the
Italian M.I.U.R. under the research project {\em Geometry of
Integrable Systems.}

\section{The search for separation coordinates}

Let $\CQ$ be an $n$-dimensional manifold, that can be thought of as the 
configuration space of a mechanical system (the sphere, for the
\neusys), and let $H$ be a function on $T^*\CQ$ (the
Hamiltonian). In this section we will present a method to look
for the variables in which the \HJ\ equation for $H$ separates.
This strategy will prove to be quite efficient in the case of
the \neusys. 

The first important point, due to Nijenhuis \cite{Nij}, is 
the idea of the geometrization of a coordinate system. If
$\{q^i\}$ are local \coos\ on an open subset $U\subset\CQ$,
then the tensor field $L$ of type $(1,1)$ defined by
\[
L\frac{\del}{\del q^i}=q^i \frac{\del}{\del q^i}
\]
has vanishing Nijenhuis torsion, i.e.,
\[
[LX,LY]-L[LX,Y]-L[X,LY]+L^2[X,Y]=0
\]
for every pair $(X,Y)$ of vector fields on $U$. Viceversa, 
a tensor field $L$ of type $(1,1)$ whose torsion is zero induces
local
\coos\ in a neighborhood of any point where the eigenvalues
$(\la_1,\dots,\la_n)$ of $L$ are distinct. In many cases these
eigenvalues are functionally independent, so that 
\[
L\frac{\del}{\del \la_i}=\la_i \frac{\del}{\del \la_i}\ ,
\]
i.e., they can be chosen as the
\coos\ associated with $L$. In the following we will suppose to
be in this situation, that is, we will look for a tensor field
$L$ with vanishing torsion and functionally independent
eigenvalues.
 
Once we have replaced the coordinate system with the geometric 
object 
$L$, it is quite natural to look for a ``compatibility
condition'' between $L$ and a function $H\in C^\infty(T^*\CQ)$
entailing that the coordinates given by $L$ are separation
variables for $H$. This condition can be obtained by lifting
$L$ from $\CQ$ to $T^*\CQ$ by means of a procedure called
``complete lifting'' \cite{YanoIshihara,CraCanSar}. It gives
rise to a torsionless tensor field $N$ of type $(1,1)$ on
$T^*\CQ$, described in  fibered coordinates $(q^i,p_i)$ as
\[
N\frac{\del}{\del q^i}= L_i^j\frac{\del}{\del q^j}+\left(\frac{\del L^k_j}{\del q^i}-\frac{\del L^k_i}{\del q^j}\right)p_k\frac{\del}{\del p_j}\ ,\qquad
N\frac{\del}{\del p_i}= L^i_j \frac{\del}{\del p_j}\ .
\]
Now that we have $H$ and $N$ on the same manifold $T^*\CQ$, we can obtain the separability condition we are looking for as follows. First, we use $H$ and $N$ to construct the 2-form $\om_H=d\left(N^* dH\right)$, where $N^*$ is the adjoint of $N$. Then we consider the \ham\ \vefi\ $X_H$ and the \vefi s $NX_H$, $N^2 X_H$, \dots obtained by iteration, and we suppose that the distribution $D_H$ spanned by $X_H$, $NX_H$,\dots, $N^{n-1} X_H$ is 
$n$-dimensional (i.e., that these \vefi s are linearly independent at any point). 
\begin{theorem}
In the above-mentioned hypotheses, the coordinates associated
with $L$ are separation variables for $H$ if and only if the
2-form $\om_H$ annihilates the distribution $D_H$: 
\begin{equation}
\label{sepcon}
\om_H|_{D_H}=0\ .
\end{equation}
\end{theorem}
The proof consists in writing the conditions 
\begin{equation}
\label{sepcons}
\om_H\left(N^k X_H, N^l X_H\right)=0\ ,\qquad\mbox{for $k,l=0,\dots,n-1$,} 
\end{equation}
in the canonical coordinates $(\la_1,\dots,\la_n,\mu_1,\dots,\mu_n)$,
where the $\la_i$ are the eigenvalues of $L$ and the $\mu_i$ are their
conjugate momenta. Then, taking into account that $N$ is diagonal in
these coordinates, it is not difficult to realize that the equations
\rref{sepcons} are equivalent to the classical \LC\ separability
conditions (\cite{DKN}, p.\ 208).

Therefore, we have found the {\em separability condition\/} between
$L$ and $H$. It is simply the vanishing of a suitable 2-form on the
distribution generated from $X_H$ by means of the complete lift $N$ of
$L$. We stress that this condition is a concise and, above all,
intrinsic form of the \LC\ equations. This means that one can check
the separability of a given \ham\ $H$ in the \coos\ associated with a
tensor field $L$ {\em before\/} computing these \coos. Moreover, one
can impose condition \rref{sepcon} on $L$ to search for \secoo\ for
$H$. In Section 4 we will show how to exploit this fact in order to 
systematically deduce the separability of the \neusys.

\section{The \bih\ meaning of the separability condition}

Before applying the results of the previous section to the \neusys,
let us make some comments on their geometrical meaning.

1. Let $\CQ$, $L$, and $N$ be as in Section 2. Then the
cotangent bundle of $\CQ$ is a
\varb\
\cite{IMM}.  The first \parp\ is the canonical one,
\[
\{F,G\}=\om(X_F,X_G)\ ,
\]
while the second one is given by
\[
\{F,G\}'=\om(NX_F,X_G)\ .
\]
If $P$ (resp.\ $P'$) is the Poisson tensor of $\parpu$ (resp.\
$\parpu'$), this means that $P'=NP$.

2. Assume that $H$ fulfills
the separability conditions
\rref{sepcon}. Then the  distribution $D_H$ is integrable, so
that there (locally) exist $n$ independent functions
$H_1,\dots,H_n$ which are constant on the leaves of $D_H$. They
are (local) first integrals for $X_H$. Notice however that the
\ham\ vector field associated with the separable \ham\ $H$ is
not (in general, and in the particular example of the \neusys)
\bih. 

3. The distribution $D_H$ is \bil, that is,
\[
\{H_i,H_j\}=0\ ,\qquad \{H_i,H_j\}'=0\qquad\mbox{for all $i,j$.}
\]
As explained in \cite{FMP2,FP}, this condition characterizes the
(integrable) \ham\ systems that are separable in the coordinates
induced by $L$. We remark in passing that \bil\ foliations play an
important role 
in the study of special K\"ahler manifolds \cite{Hitchin}.

4. As we have already said in the previous section, the \secoo\
$(\la_i,\mu_i)$ are canonical for $\om$ (this is obvious) and for $N$,
\[
N\frac{\del}{\del \la_i}=\la_i \frac{\del}{\del \la_i}\ ,\qquad
N\frac{\del}{\del \mu_i}=\la_i \frac{\del}{\del \mu_i}\ .
\]
They are often called Darboux-Nijenhuis \coos\ (since $N$ 
is sometimes called the Nijenhuis tensor) and have been used as
\secoo\ in, e.g., \cite{mt97,Bl98,FMPZ2,fmt,ybzeng}.

\section{The case of the Neumann system}

In this section we will exploit the separability condition
\rref{sepcon} to find the variables of separation
for the \neusys. To this aim, we will look for a torsionless
tensor field
$L$ of type $(1,1)$ on $S^2$ such that its
complete lift $N$ on $T^*S^2$ satisfies
\begin{equation}
\label{sepcon2}
\om_H(X_H,NX_H)=0\ ,
\end{equation}
where $\om_H=d(N^*dH)$ and $H$ is the Neumann \ham. Due to the form of
the constraint and of the potential, it is quite natural to perform
the computations in the local coordinates $X=x^2$, $Y=y^2$,
parametrizing every connected component of the
subset obtained by removing from $S^2$ its intersections
with the coordinate planes. In these
coodinates the  Hamiltonian of the Neumann system has the form:
\begin{equation}
  \label{hamneu}
\begin{split}
  H= &\, 2\left[X(1-X) p_X^2-2 XY p_Xp_Y+Y(1-Y) p_Y^2\right]\\
     &  +\frac12 (a_1-a_3)X+\frac12(a_2-a_3)Y\ ,
\end{split}
\end{equation}
where $(p_X,p_Y)$ are the conjugate momenta of $(X,Y)$. 
The unknown tensor field $L$ can be written as 
\begin{equation}
  \label{lneu}
\begin{split}
  &L^*(dX)=A dX+B dY\\
&L^*(dY)=C dX+D dY\ ,
\end{split}
\end{equation}
where $(A,B,C,D)$ are functions of the coordinates
$(X,Y)$ that must satisfy two additional conditions. The first one
is that the torsion of $L$ has to vanish, that is,
\[
\begin{split}
&B\left({C}_{Y}-{D}_{X}\right)-{B}{A}_{X}-{D}{A}_{Y}+
{A}{A}_{Y}+{C}{B}_{Y}=0\\ 
&C\left({B}_{X}-{A}_{Y}\right)-{A}{D}_{X}-{C}{D}_{Y}+
{B}{C}_{X}+{D}{D}_{X}=0\ .
\end{split}
\]
The second one is the independency of the eigenvalues of $L$. Our aim is 
to find $(A,B,C,D)$ in such a way to verify also the separability
conditions \rref{sepcon2}. Thus we need to compute the complete lift $N$
of $L$, which turns out to be given by
\begin{equation}
  \label{eq:1.4.5}
  \begin{split}
N^* d X&=A dX+BdY\\
N^* d Y&=C dX+DdY\\
N^* d p_X&=-\left[p_X\left({B}_{X}-{A}_{Y}\right)+
p_Y\left({D}_{X}-{C}_{Y}\right)\right] dY+A dp_X+Cdp_Y\\
N^* d p_Y&=\left[p_X\left({B}_{X}-{A}_{Y}\right)+
p_Y\left({D}_{X}-{C}_{Y}\right)\right] dX+B dp_X+Ddp_Y\ .
\end{split}
\end{equation}
Then \rref{sepcon2} becomes a differential equation where the coordinates $(X,Y)$, the momenta $(p_X,p_Y)$, and the
unknown functions $(A,B,C,D)$ and their derivatives appear polynomially. This
suggests to seek for a solution which also depends polynomially on the
coordinates. Trying the simplest solution, one finds
\begin{equation}
  \label{eq:1.4.9}
  \begin{split}
A&=(a_3-a_1)X+a_1\\
B&=(a_3-a_2)X\\
C&=(a_3-a_1)Y\\
D&=(a_3-a_2)Y+a_2\ ,
\end{split}
\end{equation}
which leads to the tensor field $L$ defined by
\begin{equation}
  \label{eq:1.4.10}
  \begin{split}
L^*(dX)&=a_1 dX+X d[(a_3-a_1)X+(a_3-a_2)Y]\\
L^*(dY)&=a_2 dY+Y d[(a_3-a_1)X+(a_3-a_2)Y]\ .
\end{split}
\end{equation}
It can be checked that $L$ can be extended to the whole sphere. Indeed, it is the conformal Killing tensor associated with the spheroconical coordinates \cite{Ben}. They are the eigenvalues $(\la_1,\la_2)$ of $L$, since
\[
  \begin{aligned}
\det(\la I-L)
&=\la^2-\la [(a_3-a_1)X+(a_3-a_2)Y+(a_1+a_2)]\\
&\qquad + a_2(a_3-a_1)X+a_1(a_3-a_2)Y+a_1a_2\\
&=\la^2(x^2+y^2+z^2) \\
&\qquad -\la[(a_3-a_1)x^2+(a_3-a_2)y^2+(a_1+a_2)(x^2+y^2+z^2)]\\
&\qquad +a_2(a_3-a_1)x^2+a_1(a_3-a_2)y^2+a_1a_2(x^2+y^2+z^2)\\
&=\la^2(x^2+y^2+z^2)-\la[(a_2+a_3)x^2+(a_1+a_3)y^2+(a_1+a_2)z^2]\\
&\qquad + [a_2a_3 x^2+a_1a_3y^2+a_1a_2z^2]\\
&=(\la-a_2)(\la-a_3)x^2+(\la-a_1)(\la-a_3)y^2+(\la-a_1)(\la-a_2)z^2\\
&=\left[\frac{x^2}{\la-a_1}+\frac{y^2}{\la-a_2}+\frac{z^2}{\la-a_3}
\right](\la-a_1)(\la-a_2)(\la-a_3)\>.
\end{aligned}
\]
Hence,
\[
\frac{(\la-\la_1)(\la-\la_2)}{(\la-a_1)(\la-a_2)(\la-a_3)}=
  \frac{x^2}{\la-a_1}+\frac{y^2}{\la-a_2}+\frac{z^2}{\la-a_3}\ ,
\]
meaning that the eigenvalues of $L$ are the spheroconical
cooordinates on $S^2$. We have thus deduced the usual separation
coordinates of the Neumann system simply using the separability
condition \rref{sepcon} discussed in Section 2.

\section{Integrals of motion for the Neumann system}

The aim of this section is to show that in our setting one can easily
describe other interesting features of the \neusys, such as a simple
construction of the integrals of motion and the existence of a \bih\
formulation in a suitable extension of the phase space. The crucial
point is that the \ham\ $H$ of the \neusys\ and the
complete lift $N$ of the tensor field \rref{eq:1.4.10} satisfy
the stronger condition 
\begin{equation}
\label{sepconstr}
d(N^*dH)=dp_1\wedge dH\ ,\qquad \mbox{where
$p_1=\mbox{tr}\,L=\la_1+\la_2$,}
\end{equation}
implying the separability condition \rref{sepcon}. This fact has
important consequences, that we are going to show in the general
setting considered in Sections 2 and 3.

Let $L$ be a torsionless tensor field of type $(1,1)$,
with functionally independent eigenvalues, on an
$n$-dimensional manifold $\CQ$, so that we can endow the
symplectic manifold
$(T^*\CQ,\parpu)$ with the additional \parp
\[
\{F,G\}'=\om(NX_F,X_G)\ ,
\]
using the complete lift $N$ of $L$. Let 
\[
\det(\la I-L)=\la^n-\left(p_1\la^{n-1}+p_2\la^{n-2}+\dots+p_n\right)
\]
be the characteristic polynomial of $L$, and suppose that $H\in
C^\infty(T^*\CQ)$ satisfies
\begin{equation}
\label{sepstr}
d(N^*dH)=dp_1\wedge dH\ .
\end{equation}
Then one can easily show that $\om_H=d(N^*dH)$ vanishes on
the distribution $D_H$, so that the separability condition
\rref{sepcon} is fulfilled and $H$ is separable in the coordinates
associated with $L$. We know from Section 3 that there are (local)
integrals of motion for $X_H$. If \rref{sepstr} holds, they can be
easily found by writing it in the form $d(N^*dH-p_1 dH)=0$, and
choosing $H_2$ such that $dH_2=N^*dH-p_1 dH$. Then (as shown in
\cite{IMM} in the case where $H$ is quadratic in the momenta) also the 1-form
$N^*dH_2-p_2 dH$ is closed, and the process can be iterated, so that
the integrals of motion for $X_H$ can be determined from the recursion
relations
\begin{equation}
\label{froch}
\begin{split}
& dH_2=N^*dH-p_1 dH\\
& dH_3=N^*dH_2-p_2 dH\\
& \phantom{dH_3\ }\vdots \\
& dH_n=N^*dH_{n-1}-p_{n-1} dH\ .
\end{split}
\end{equation}
Moreover, we have that 
\begin{equation}
\label{frochn}
0=N^*dH_{n}-p_{n} dH\ .
\end{equation}
For the Neumann system the global existence of the function $H_2$ is
garanteed by the fact that $T^*S^2$ is simply connected. One finds
that, in local coordinates, 
\[
\begin{aligned}
H_2 &=2(a_1 {p_Y}^2 Y+a_2 {p_X}^2 X)(X+Y-1)-2a_3
XY(p_Y-p_X)^2\\
&\ -\frac12 a_2(a_1-a_3) X-\frac12 a_1(a_2-a_3) Y\ ,
\end{aligned}
\]
which coincides with the constant of motion provided by the Lax
matrix. Notice that 
the \ham\ \vefi\ $X_H$ is not \bih\ on $T^*\CQ$. Nevertheless, from
\rref{frochn} we can conclude that 
\[
X_H=\{\cdot,H\}=\frac1{p_n}\{\cdot,H_n\}'\ .
\]
In the terminology of \cite{Brouzet}, this means that $X_H$ is a
quasi-bi-Hamiltonian \vefi\ (see also \cite{Bl98,mt97}). Next we
will show that we can obtain a \bih\ representation of $X_H$ on
the extended phase space $\CM:=T^*\CQ\times\mathbb R$, following
\cite{IMM,CraSarTho}. The first step is the extension of the \bih\
structure from $T^*\CQ$ to $\CM$. If $c\in\mathbb R$ is a coordinate
in the ``additional dimension'' and $F\in C^\infty(T^*\CQ)$, then the
new \parp s are defined as
\begin{eqnarray*}
\{F,c\}_\CM &=& 0\\
\{F,c\}'_\CM &=& \{F,H\}-c\{F,p_1\}\ .
\end{eqnarray*}
They endow $\CM$ with a \bih\ structure. The second step is to notice
that the recursion relations \rref{froch} on the $H_i$ become the usual
Lenard relations on the functions ${\hat H}_i$ defined on $\CM$ as 
\[
{\hat H}_0 = c\ ,\qquad 
{\hat H}_i = H-cp_i\quad \mbox{for $i=1,\dots,n$.}
\]
Indeed, we have 
\[
\begin{array}{rcl}
\{{\hat H}_0,\cdot\}_\CM &=& 0\\
\{{\hat H}_1,\cdot\}_\CM &=& \{{\hat H}_0,\cdot\}'_\CM\\
 & \vdots & \\
\{{\hat H}_n,\cdot\}_\CM &=& \{{\hat H}_{n-1},\cdot\}'_\CM\\
 0 &=& \{{\hat H}_{n},\cdot\}'_\CM\ .
\end{array}
\]
In the Neumann case, the restriction to $c=0$ of
the \bih\ \vefi\ $\{{\hat H}_1,\cdot\}_\CM = \{{\hat
H}_0,\cdot\}'_\CM$ is the Neumann \vefi. Thus we have shown
that the \neusys\ admits a \bih\ formulation in an extended
phase space, recovering in this way a result of \cite{Bl98}. 

\section{Concluding remarks}

1. The extension
process of a chain of the form \rref{froch} into a Lenard chain is the
opposite of the reduction technique presented in \cite{FMPZ2,creta}, where
a suitable quotienting produces a chain like \rref{froch} from a Lenard chain.
\par\noindent
2. All the results presented here can be extended to the
$n$-dimensional \neusys. It would be interesting to compare our
approach with the one based on the Lax representation.
\par\noindent
3. The condition \rref{sepconstr} appears also in \cite{CraSarTho},
in the case where $(\CQ,g)$ is a Riemannian manifold and $H=\frac12
g^{ij}(q)p_ip_j+V(q)$. The authors show that \rref{sepconstr} implies
that $L$ is a conformal Killing tensor of $g$ with vanishing
torsion. Notice however that in our approach to separability the
Riemannian structure of $\CQ$ plays no role and the important objects
are \bil\ foliations on \varb s. 
\par\noindent
4. The idea of extending the phase space in order to obtain a \bih\
formulation of a given (\ham) \vefi\ goes back, to the best of our
knowledge, to \cite{SRW}, where it has been used also in the context
of separation of variables.

\end{document}